\documentclass[final,3p,times,twocolumn]{elsarticle}

\usepackage[T1]{fontenc}
\usepackage[utf8]{inputenc}
\usepackage{lmodern}
\usepackage{hyperref}
\usepackage{lineno}
\usepackage{subfig}
\usepackage{graphicx}
\usepackage{threeparttable}
\usepackage{booktabs}
\usepackage[section]{placeins}
\usepackage{pdflscape}
\usepackage{afterpage}
\usepackage{xcolor}
\usepackage{ulem}

\usepackage{amsmath}
\usepackage[version=4]{mhchem}
\usepackage[table-align-text-post=false, range-units=single, range-phrase=\textendash, separate-uncertainty=false, multi-part-units=single, per-mode=symbol-or-fraction]{siunitx}

\newcommand{\tnotes}[1]{{\vspace{0.02cm}\par\leftskip0.15cm \rightskip\leftskip\scriptsize\linespread{0.5}#1\newline\vspace{-\baselineskip}\par}}
\newcommand{\revchanges}[1]{{#1}}

\definecolor{accent}{HTML}{FF0266}
\definecolor{darkblue}{HTML}{0336FF}

\journal{Journal of Molecular Spectroscopy}

\begin{document}
\begin{frontmatter}
\title{Extending the rotational spectrum of cyclopentadiene towards higher frequencies and vibrational states}

\author[cologne]{Luis Bonah}
\author[cologne]{Benedikt Helmstaedter}
\author[rennes]{Jean-Claude Guillemin}
\author[cologne]{Stephan Schlemmer}
\author[cologne]{Sven Thorwirth}

\affiliation[cologne]{
    organization={I.\ Physikalisches Institut, Universität zu Köln},
    addressline={Zülpicher Str.\ 77}, 
    city={Köln},
    postcode={50937},
    country={Germany}
}

\affiliation[rennes]{
            organization={Univ Rennes, Ecole Nationale Supérieure de Chimie de Rennes},
            addressline={ISCR-UMR 6226}, 
            city={Rennes},
            postcode={35000}, 
            country={France}
}

\begin{abstract}
Cyclopentadiene (\ce{\textit{c}-C5H6}) is a cyclic pure hydrocarbon that was already detected astronomically towards the prototypical dark cloud TMC-1 (Cernicharo et al. 2021, \textit{Astron. Astrophys.} \textbf{649}, L15).
However, accurate predictions of its rotational spectrum are still limited to the microwave region and narrow quantum number ranges.
In the present study, the pure rotational spectrum of 
cyclopentadiene was measured in the frequency ranges \SIrange{170}{250}{GHz} and \SIrange{340}{510}{GHz} to improve the number of ground vibrational state assignments by more than a factor of 20, resulting in more accurate rotational parameters and the determination of higher-order centrifugal distortion parameters.
Additionally,  %
vibrational satellite spectra of
cyclopentadiene in its eight energetically lowest vibrationally excited states were analyzed for the first time.
Coriolis interactions between selected vibrational states were identified and treated successfully in combined fits. Previous microwave work on the three singly \ce{^{13}C} substituted isotopologues was extended significantly also covering frequency ranges up to 250\,GHz.
The new data sets permit reliable frequency predictions for the isotopologues and vibrational satellite spectra far into the sub-mm-wave range.
\revchanges{
Finally, the experimental rotational constants of all available isotopologues and calculated zero-point vibrational contributions to the rotational constants were used to derive a semi-experimental equilibrium structure of this fundamental ring molecule.
}
\end{abstract}


\end{frontmatter}

\section{Introduction}
\label{sec:Introduction}
Cyclopentadiene (\ce{\textit{c}-C5H6}) is a polar cyclic hydrocarbon with an appreciable dipole moment of 0.419(4)\,D \cite{Scharpen1965}.
In the laboratory, its rotational spectrum was first studied in the microwave by Laurie in 1956~\cite{Laurie1956}.
Later studies measured more microwave transitions~\cite{Scharpen1965,Flygare1970} and extended the frequency coverage into the mm-wave region up to \SI{390}{GHz}~\cite{Bogey1988}.
\begin{figure}[tb]
    \centering
    \includegraphics[width=\linewidth]{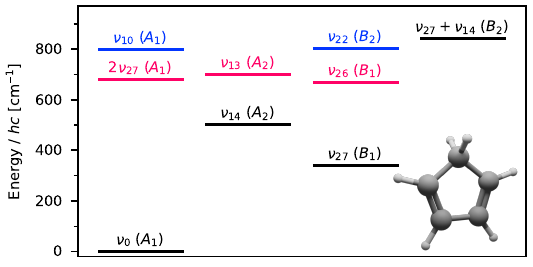}
    \caption{The vibrational states of cyclopentadiene below \SI{850}{cm^{-1}}. The respective symmetries are given in parentheses.
    The two interaction systems are indicated by blue and red color.}
    \label{fig:EnergyLevels}
\end{figure}
However, even in the most recent work by Bogey et al.~\cite{Bogey1988} only 99 lines were reported. %
The resulting parameters have high uncertainties and no sextic or higher-order parameters were fitted.
Therefore, the accuracy for predictions outside the measured frequency and quantum number range leaves much room for improvement. %
In addition to the main isotopologue, the three singly \ce{^{13}C} substituted isotopologues and different deuterium-substituted isotopologues were examined~\cite{Damiani1976, Scharpen1965} in the microwave regime, but no mm-wave data are available.
No vibrational satellite spectra have been reported so far for any isotopic species and only 
one vibrational band has been studied at high spectral resolution, 
the $\nu_{26}$ fundamental mode of the parent isotopologue~\cite{Boardman1990}.
However, eight vibrational states lie below \SI{850}{cm^{-1}} with the methylene-rocking motion being the lowest at about \SI{350}{cm^{-1}} (see \autoref{fig:EnergyLevels})~\cite{Castellucci1975}.
The narrow spacing between some vibrational levels suggests they might be subject to considerable interactions.

Cyclopentadiene was detected astronomically towards TMC-1~\cite{Cernicharo2021} via five lines with low quantum numbers ($2 \leq J \leq 5$ and $K_a \leq 3$) around \SI{38}{GHz} and \SI{46}{GHz}, all of which belong to the ground vibrational state of the main isotopologue.

This work aims at extending the frequency and quantum number coverage of the main isotopologue and the singly \ce{^{13}C} substituted isotopologues.
Additionally, for the parent isotopic species, rotational analyses of all vibrationally excited states below \SI{850}{cm^{-1}} are presented, including a treatment of two resonance systems identified for the first time.

\section{Experimental details}
\label{sec:Experimental Details}
High-resolution broadband spectra of cyclopentadiene were recorded with an absorption spectrometer in the frequency ranges \SIrange{170}{250}{GHz} and \SIrange{340}{510}{GHz}.
A commercially available radiation source from Virginia Diodes was combined with two different amplifier-multiplier chains to reach the desired frequency ranges.
A \SI{5}{m} absorption cell was used in a double-pass setup for a total absorption path of \SI{10}{m}.
On the detection side, a Schottky detector was employed with subsequent preamplifiers and a lock-in amplifier.
The radio frequency was frequency modulated by the synthesizer and a \textit{2f} demodulation of the detector signal was performed by the lock-in amplifier to increase the signal-to-noise ratio (SNR).
The resulting lineshapes look similar to a second derivative Voigt profile.
The spectrometer was described in greater detail elsewhere previously~\cite{MartinDrumel2015}.

Commercially, cyclopentadiene is not available in monomeric form but as a Diels-Alder adduct, dicyclopentadiene. It may however
be produced from the adduct through a thermally induced retro-Diels-Alder reaction.  
In the present study, the thermolysis used a quartz tube, an oven with a \SI{10}{cm} heating zone and a temperature of about \SI{560}{\celsius} which was determined to yield optimal production for this specific setup~\cite{Helmstaedter2024}.
A needle valve attached to the sample container allowed to precisely set the precursor flow.
To increase the vapor pressure of dicyclopentadiene and prevent clogging of the needle valve, the sample container and the valve were resistively heated to about \SI{55}{\degree C}.
The rotational temperature in the absorption cell was room temperature as is apparent from the rotational spectrum (see e.g. the good agreement between the experimental intensities and the predictions performed at \SI{300}{K} in \autoref{fig:TypicalPattern}).
Due to the thermolysis, all measurements were performed under mild flow conditions while keeping the pressure in the cell in the range of \SIrange{20}{30}{\micro bar}.
Standing waves were removed from the raw spectral data via Fourier filtering with a self-written script\footnote{\texttt{\url{https://github.com/Ltotheois/SnippetsForSpectroscopy/tree/main/FFTCorrection}}}.

\section{Spectroscopic fingerprint of cyclopentadiene}
\label{sec:Analysis}

\begin{figure*}[tb]
    \centering
    \includegraphics[width=\linewidth]{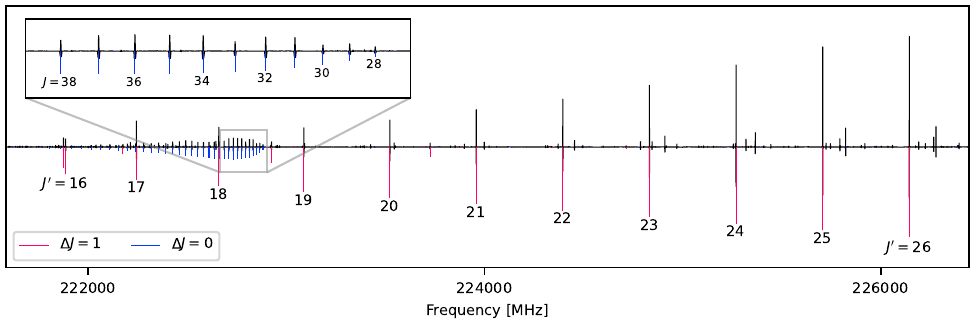}
    \caption{Part of the experimental spectrum of cyclopentadiene in black. Typical $\Delta J=1$ and $\Delta J=0$ patterns of the $b$-type spectrum of cyclopentadiene are highlighted by the corresponding predictions in red and blue, respectively.
    The $R$-branch transitions are governed by $\Delta J_{\Delta K_a, \Delta K_c} = 1_{\pm 1, 1}$ with $K'_a = 26 - J'$ or $27 - J'$ and $J'$ as indicated.
    In the zoom-in, $Q$-branch transitions obey $\Delta J_{\Delta K_a, \Delta K_c} = 0_{1, -1}$ and $K'_a = J-27$ or $J - 26$.} %
    \label{fig:TypicalPattern}
\end{figure*}

Cyclopentadiene is an asymmetric top rotor close to the oblate limit as seen from its Ray's asymmetry parameter $\kappa = (2B-A-C)/(A-C) = \SI{0.90}{}$.
The main isotopologue has $C_{2v}$ symmetry\footnote{The naming of the symmetry species $B_1$ and $B_2$ is swapped in this work compared to some older works, e.g. Castellucci et al.~\cite{Castellucci1975}.} resulting in the two spin species ortho ($K_a + K_c$ odd) and para ($K_a + K_c$ even) with statistical weights of 9 and 7, respectively~\cite{Boardman1990, Cernicharo2021}.
Only the $b$-type dipole moment is non-zero (\SI{0.419(4)}{D}~\cite{Scharpen1965})
and the rotational spectrum
mainly comprises two characteristic patterns for $\Delta J = 1$ ($R$-branch) and $\Delta J = 0$ ($Q$-branch) transitions which are highlighted in \autoref{fig:TypicalPattern}.

The $R$-branch line series
consist of $\Delta J_{\Delta K_a, \Delta K_c} = 1_{\pm 1, 1}$ transitions. For high $J$ values the $\Delta K_a = +1$ transitions are blended with the respective $\Delta K_a = -1$ transitions.
Transitions with the same $J+K_a$ value form patterns (see red lines in \autoref{fig:TypicalPattern}) that increase in $J$ with increasing frequency. When $K_a$ is far from $J$, the lines are spaced almost equidistantly (about \SI{430}{MHz}).
These patterns repeat every $\SI{8.54}{GHz}$ corresponding to $2 (B+C-A) = 2C$ for the limiting case of an oblate symmetric top.

$Q$-branches comprise $\Delta J_{\Delta K_a, \Delta K_c} = 0_{1, -1}$ transitions and again the two asymmetry components are typically blended.
Lines belonging to a single pattern (see blue lines in \autoref{fig:TypicalPattern}) share the same $K_c$ value.
Going from high to low frequencies, the pattern starts with the
$K''_a=0$ transition and its respective blended $K''_a=1$ transition.
Subsequent lines (at lower frequencies) increase in $J$ and $K_a$.
Similarly, the distance between lines increases slightly when $J$ values increase.
These $Q$-type patterns repeat about every $\SI{8.1}{GHz}$, which for the limiting case of an oblate symmetric top would be given by $2 (2B - A - C) = 2(B-C)$.

\begin{table}[tb]
\caption{The resulting rotational parameters for the ground vibrational states of the main isotopologue and the three singly \ce{^{13}C} substituted isotopologues.}
\label{tab:isotopologues}
\begin{threeparttable}
\resizebox{\linewidth}{!}{

\centering
\begin{tabular}{l l *{ 4 }{S[table-format=4.9]}}
\toprule
\multicolumn{2}{l}{Parameter} & \ce{$c$-C5H6} & \text{$\ce{1-^{13}C}$} & \text{$\ce{2-^{13}C}$} & \text{$\ce{3-^{13}C}$} \\
\midrule
$ A                    $&/$ \si{\mega\hertz}     $&     8426.108825(35)&       8226.0534(18)&      8420.04351(98)&      8345.13300(94)\\ 
$ B                    $&/$ \si{\mega\hertz}     $&     8225.640352(33)&       8219.4832(18)&       8040.4326(14)&       8108.7105(12)\\ 
$ C                    $&/$ \si{\mega\hertz}     $&     4271.437296(30)&      4217.75907(40)&      4219.40981(37)&      4219.06493(33)\\ 
$ -D_{J}               $&/$ \si{\kilo\hertz}     $&       -2.692726(21)&         -2.6502(22)&         -2.6318(19)&         -2.6357(17)\\ 
$ -D_{JK}              $&/$ \si{\kilo\hertz}     $&        4.059634(34)&          3.9985(49)&          3.9703(41)&          3.9771(38)\\ 
$ -D_{K}               $&/$ \si{\kilo\hertz}     $&       -1.682765(22)&         -1.6585(26)&         -1.6472(22)&         -1.6498(21)\\ 
$ d_{1}                $&/$ \si{\hertz}          $&         -42.220(10)&\text{a}&          -17.2(1.3)&          -12.7(1.3)\\ 
$ d_{2}                $&/$ \si{\hertz}          $&         -0.6014(36)&\text{a}&           -5.06(60)&            1.90(51)\\ 
$ H_{J}                $&/$ \si{\milli\hertz}    $&          1.0208(37)&\text{a}&\text{a}&\text{a}\\ 
$ H_{JK}               $&/$ \si{\milli\hertz}    $&         -4.0297(71)&\text{a}&\text{a}&\text{a}\\ 
$ H_{KJ}               $&/$ \si{\milli\hertz}    $&          5.0425(73)&\text{a}&\text{a}&\text{a}\\ 
$ H_{K}                $&/$ \si{\milli\hertz}    $&         -2.0316(41)&\text{a}&\text{a}&\text{a} \\
\midrule
\multicolumn{2}{l}{Transitions}         &   3510    &   228   &  228    &  235    \\
\multicolumn{2}{l}{Lines}               &   1992    &   120   &  143    &  145    \\
\textit{RMS} &/\si{\kilo\hertz}         & 21.40     & 14.97   & 16.72   & 16.31  \\ 
\multicolumn{2}{l}{\textit{WRMS}}       & 0.79      & 0.44    & 0.49    & 0.46   \\ 
\bottomrule
\end{tabular}}
\tnotes{
Fits performed with SPFIT in the S-reduction and $\text{III}^\text{l}$ representation.
Standard errors are given in parentheses.
Parameters of the ground vibrational state are applied to all vibrationally excited states and difference values are fitted.
$^a$~Parameter was fixed to the global value.
}
\end{threeparttable}
\end{table}

Spectroscopic assignment was carried out via Loomis-Wood plots as implemented in the LLWP software~\cite{Bonah2022}.
Pickett’s SPFIT was used for least squares fitting to an asymmetric top Hamiltonian in the 
S-reduction and $\text{III}^\text{l}$ representation~\cite{Pickett1991}.
Line uncertainties were determined with a semi-automatic procedure described previously~\cite{Bonah2024}.

\revchanges{Complementary quantum chemical calculations in support of the spectroscopic analysis  
were performed using the CFOUR suite of programs \citep{Matthews2020,Harding2008} 
using strategies outlined elsewhere \cite{Puzzarini2010}.
Most importantly, the rotation-vibration interaction constants $\alpha_i^{A,B,C}$ as well as vibrational wavenumbers \text{$\tilde{\nu_i}$} were derived from an anharmonic force field computed at the coupled-cluster (CC) singles and doubles level extended by a perturbative correction for the contribution
from triple excitations, CCSD(T),~\cite{Raghavachari1989} in combination with the ANO0 basis set \cite{Almlf1987} and in the frozen core (fc) approximation. A high-level equilibrium structure of cyclopentadiene was calculated at the CCSD(T)/cc-pwCVQZ level (cf. Ref \cite{thorwirth_JMS_350_10_2018} and references therein) considering all electrons (ae) in the correlation treatment.}

Assigning the spectrum of the main isotopologue was straightforward due to literature data covering frequencies up to \SI{390}{GHz}~\cite{Bogey1988} and readily visible patterns.
Literature data from frequency ranges not covered here~\cite{Scharpen1965,Flygare1970, Bogey1988} were incorporated into our analysis except data from Laurie~\cite{Laurie1956} as their data shows a systematic offset (mean $\nu_\text{obs} - \nu_\text{calc}$ value of \SI{136(114)}{kHz}) and because most of their lines were remeasured in later works.

Substituting a single \ce{^{12}C} with a \ce{^{13}C} atom has considerable influence on the observed spectrum.
When substituting the \ce{1-C} atom\footnote{The labeling of carbon atoms is equivalent to the one in Scharpen et al.~\cite{Scharpen1965}, where \ce{1-C} is the methylene carbon atom and the other carbons are numbered sequentially.}, the $a$- and $b$-axes are swapped.
\begin{figure*}[tbh]
    \centering
    \includegraphics[width=1\linewidth]{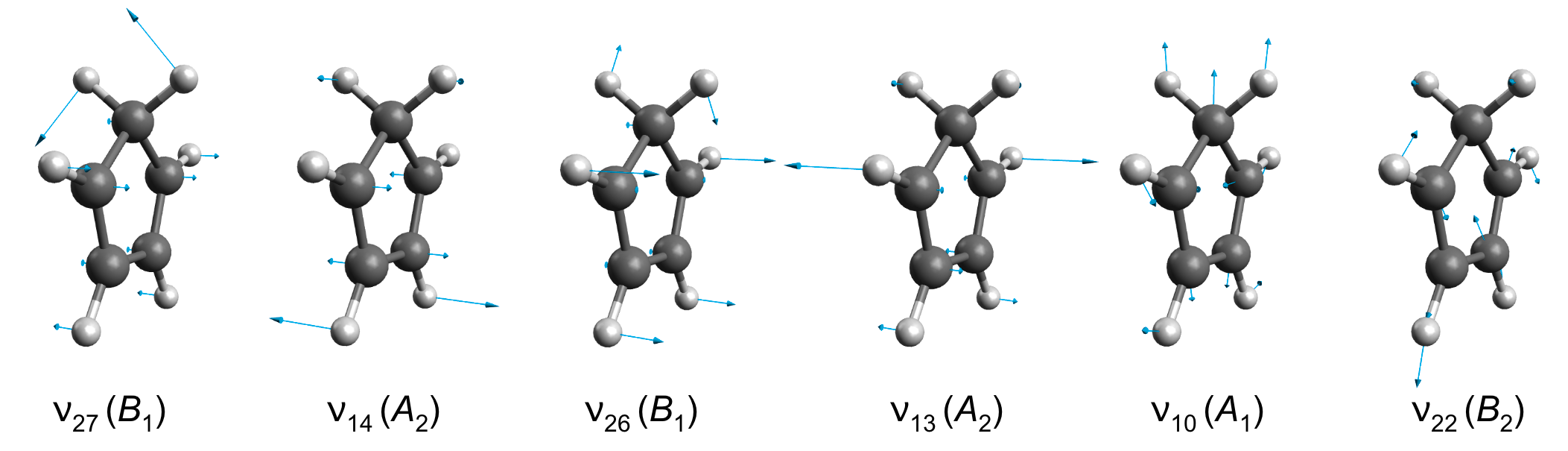}
    \caption{The six energetically lowest vibrational fundamentals sorted to increasing energy from left to right.}
    \label{fig:VibrationModes}
\end{figure*}
Therefore, the $A$ constant of the \ce{1-^{13}C} isotopologue should be compared with the $B$ constants of the other isotopologues and an $a$-type spectrum is observed.
Substituting the \ce{2-C} or \ce{3-C} atom lifts the $C_{2v}$ symmetry and rotates the principal axes in the $a-b$ plane introducing an $a$-type dipole moment with a ratio $\mu_a/\mu_b$ of \SI{0.12}{} and \SI{0.50}, respectively~\cite{Scharpen1965}.
Literature data covering lower frequencies were included in the analyses presented here~\cite{Scharpen1965}.
Due to the low natural abundance, the analysis of the singly \ce{^{13}C} substituted isotopologues was limited by their low SNR to frequencies below \SI{250}{GHz}.

The resulting parameters for the ground vibrational states of the isotopologues (see \autoref{tab:isotopologues}) show good agreement with literature data (see \autoref{tab:LiteratureData}\footnote{The CDMS results are used for the main isotopologue as they include more lines than Bogey et al.~\cite{Bogey1988} and the $\text{III}^\text{l}$ representation is used.}), although generally not within quoted uncertainties.
\revchanges{
As no literature data were available for the quartic centrifugal distortion constants of the singly substituted \ce{^{13}C} isotopologues, these values were derived (for all analyzed isotopologues) from fc-CCSD(T)/ANO0 force field calculations (see \autoref{tab:QCCQuartics}).
The good agreement between the calculated and experimental values confirms the physical meaningfulness of the experimental parameters with the exception of the $d_2$ parameter of the parent isotopic species.
This deviation might be a consequence of the very small magnitude of this parameter for the parent species combined with the influence of zero-point vibration. Trial fits of the parent isotopic species using the $\text{I}^\text{r}$ representation reveal proper agreement between the experimental and calculated quartic centrifugal distortion constants at the cost of an overall degraded fit quality.
}
The new global fits are extended significantly regarding the number of assigned transitions as well as the frequency and quantum number coverage.
Compared to previous results, additional higher-order parameters were added and the accuracy of existing parameters was improved.

Vibrational motions of the six energetically lowest vibrational fundamentals are shown in \autoref{fig:VibrationModes}.
While the calculated vibrational wavenumbers show good agreement with literature values~\cite{Castellucci1975}, there are substantial deviations of opposite sign for the $\alpha_i^C$ values of the $\nu_{10}$ and $\nu_{22}$ fundamentals.
The magnitude combined with the mirror-like appearance of the deviations suggests that they result from contributions of Coriolis coupling to the rotational constants~\cite{Mills1972, Torneng1980, Endres2021}.
As a result, the search for the vibrational states $\nu_{10}$ and $\nu_{22}$ did not rely on calculated predictions but on finding their spectroscopic patterns by visual means alone, a task
challenged by the diminished intensity of the vibrational satellites due to their small
Boltzmann factors of only \SI{2}{\percent}.
To support the identification of low-intensity patterns, already assigned states were removed from the spectrum by deleting the experimental data points around their peak position.
This was done in an iterative process for the strongest remaining pattern(s) as shown in \autoref{fig:ReducedSpectrum}.
The effectiveness of Loomis-Wood plots was greatly improved by this procedure and allowed to assign the vibrational satellite spectra of $\nu_{10}$ and $\nu_{22}$ despite the large discrepancy between the calculated and experimental values of their $C$ rotational constants.
While molecular parameters of the eight energetically lowest vibrationally excited states of cyclopentadiene are presented here, satellites from states even higher in energy were easily visible and could be assigned readily.
However, a tentative analysis hinted towards strong interactions between multiple of these states, putting their quantitative analysis beyond the scope of this work.
\begin{landscape}
\thispagestyle{empty}
\begin{figure}[p]
    \centering
    \includegraphics[width=1\linewidth]{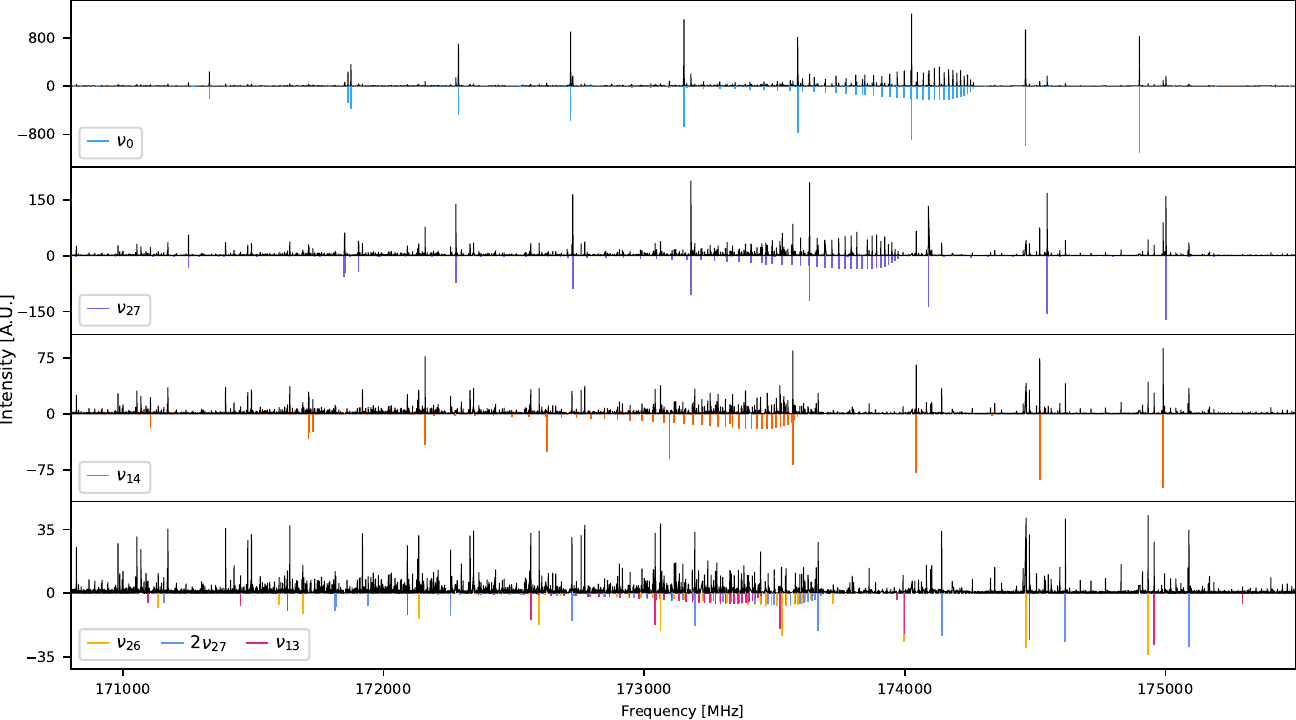}
    \caption{The process of removing known vibrationally excited states from the experimental spectrum visualized.
    In the top row, the complete spectrum is shown together with the predicted stick spectrum for the ground vibrational state in blue.
    Removing all data points within \SI{2.5}{MHz} of the assigned ground state positions yields the spectrum in the row below \revchanges{(with the $y$-axis being rescaled to match the reduced data)}. There the predicted stick spectrum of the energetically lowest vibrationally excited state $\nu_{27}$ is shown in purple.
    Similarly, rows three and four show the spectrum after additionally removing lines belonging to $\nu_{27}$ and $\nu_{14}$, respectively.
    This process greatly facilitates visual pattern recognition for vibrational states (and/or isotopologues) with lower intensities.
    The two lowest rows show that the method also has drawbacks, as some prominent lines from the predictions are missing in the experimental spectrum.
    This results from these lines being blended with transitions from already removed states.
    Combining this approach with Loomis-Wood plots increases the fault tolerance drastically.
    }
    \label{fig:ReducedSpectrum}
\end{figure}
\end{landscape}

\begin{figure}[tb]
    \centering
    \includegraphics[width=\linewidth]{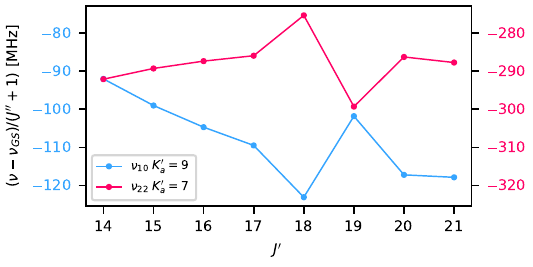}
    \caption{Resonance plot of the $K'_a= 9$ series of $\nu_{10}$ and the $K'_a=7$ series of $\nu_{22}$.
    In all transitions $\Delta J=\Delta K_a = \Delta K_c = 1$.
    The mirror image-like deviations for $J'=18$ and $19$ indicate an interaction that is centered around the respective energy levels with $J'=18$.}
    \label{fig:ResonancePlot}
\end{figure}

Within the eight energetically lowest vibrational states analyzed here, two interaction systems could be identified by mirror images in their residuals and resonance plots (see \autoref{fig:ResonancePlot}),
the $\nu_{10}$/$\nu_{22}$ dyad and the $\nu_{26}$/$2\nu_{27}$/$\nu_{13}$ triad.
The initial guess for the vibrational energy difference was approximated from the involved rotational energy levels.
\revchanges{
For the final model, 50 interstate transitions between $\nu_{13}$ and $2\nu_{27}$ as well as 20 interstate transitions between $\nu_{10}$ and $\nu_{22}$ were assigned.
They result from wavefunction mixing due to the interactions and help determine the vibrational energy difference accurately.}
The interaction parameters were systematically tested with Pyckett\footnote{Pyckett is a Python wrapper around Pickett's SPFIT and SPCAT~\cite{Pickett1991}. It can be installed via \texttt{pip install pyckett}}.
From group theory it can be derived, that Coriolis interactions occur only between vibrational states of different symmetry and for each pair of states only a single symmetry (meaning $a$-, $b$-, or $c$- symmetry) of Coriolis interactions can occur~\cite{Jahn1939, Boyd1952, Mills1965}.
\begin{table*}[tb]
\caption{The resulting rotational parameters for the eight energetically lowest vibrationally excited states. \revchanges{The two interaction systems are indicated by blue and red colored table headers similar to \autoref{fig:EnergyLevels}.} }
\label{tab:VibrationalStates}
\begin{threeparttable}
\resizebox{\linewidth}{!}{

\centering
\begin{tabular}{l l *{ 8 }{S[table-format=4.9]}}
\toprule
\multicolumn{2}{l}{Parameter} & \text{$\nu_{27}$} & \text{$\nu_{14}$} & {\color{accent} \text{$\nu_{26}$}} & {\color{accent} \text{$2\nu_{27}$}} & {\color{accent} \text{$\nu_{13}$}} & \text{$\nu_{27}+\nu_{14}$} & {\color{darkblue} \text{$\nu_{10}$}} & {\color{darkblue} \text{$\nu_{22}$}} \\
\midrule
$ A                    $&/$ \si{\mega\hertz}     $&     8429.339717(69)&     8418.480575(89)&      8408.25559(13)&       8431.9322(70)&      8395.70079(81)&        8421.515(51)&        8416.784(17)&        8444.228(20)\\ 
$ B                    $&/$ \si{\mega\hertz}     $&     8214.788334(61)&     8207.517052(80)&      8217.21858(15)&       8204.4205(78)&       8220.5890(49)&        8197.100(49)&        8227.977(17)&        8227.153(19)\\ 
$ C                    $&/$ \si{\mega\hertz}     $&     4274.157884(47)&     4274.115571(50)&     4272.629500(60)&     4276.535426(89)&     4273.326526(90)&      4276.65964(11)&        4263.061(50)&        4264.049(50)\\ 
$ -D_{J}               $&/$ \si{\kilo\hertz}     $&       -2.696942(27)&       -2.682148(34)&       -2.646176(95)&        -2.70645(22)&        -2.58322(32)&        -2.68418(41)&         -2.6101(83)&         -2.9493(84)\\ 
$ -D_{JK}              $&/$ \si{\kilo\hertz}     $&        4.062326(43)&        4.019933(63)&         3.94434(31)&         4.06999(22)&         3.72832(65)&         4.02211(92)&           4.951(12)&          3.6197(86)\\ 
$ -D_{K}               $&/$ \si{\kilo\hertz}     $&       -1.682418(30)&       -1.655183(41)&        -1.61483(26)&\text{a}&        -1.46181(34)&        -1.65897(51)&          -2.626(18)&          -1.015(15)\\ 
$ d_{1}                $&/$ \si{\hertz}          $&         -34.402(29)&         -38.531(38)&         -58.791(54)&\text{a}&          -98.7(1.7)&\text{a}&\text{a}&\text{a}\\ 
$ d_{2}                $&/$ \si{\milli\hertz}    $&              40(13)&\text{a}&\text{a}&\text{a}&\text{a}&\text{a}&\text{a}&\text{a}\\ 
\midrule
\multicolumn{2}{l}{Transitions}         &   2117    &   1736   &  1146   &  441    &  437    &  572 & 572 & 550 \\
\multicolumn{2}{l}{Lines}               &   1223    &   996    &  618    &  224    &  231     & 286 & 288 & 277 \\
\textit{RMS} &/\si{\kilo\hertz}         & 22.07     & 22.47    & 26.90\tnote{b}    & 26.90\tnote{b} & 26.90\tnote{b} & 21.79  & 25.06\tnote{c} & 25.06\tnote{c}  \\ 
\multicolumn{2}{l}{\textit{WRMS}}       & 0.85      & 0.74     & 0.86 \tnote{b}    & 0.86\tnote{b}  & 0.86\tnote{b}  &   0.60 & 0.74\tnote{c} &  0.74\tnote{c} \\ 
\bottomrule
\end{tabular}}
\tnotes{
Fits performed with SPFIT in the S-reduction and $\text{III}^\text{l}$ representation.
Standard errors are given in parentheses.
Parameters of the ground vibrational state are applied to all vibrationally excited states and difference values are fitted.
Parameters not specified were fixed to the global values, see \autoref{tab:isotopologues}.
\revchanges{Interstate transitions are counted towards the lower state for the transitions and lines statistics.}
$^a$~Parameter was fixed to the global value. $^{b,c}$~Reported values are values for the respective combined fits.
}
\end{threeparttable}
\end{table*}
\begin{table}[tb]
\caption{The resulting energy differences (colored analog to \autoref{fig:EnergyLevels} and \autoref{tab:isotopologues}) and interaction parameters for the interactions between $\nu_{10}$ and $\nu_{22}$ as well as $\nu_{26}$, $2\nu_{27}$, and $\nu_{13}$.}
    \label{tab:InteractionParameters}
\begin{threeparttable}
\centering
\resizebox{\linewidth}{!}{
\begin{tabular}{c c c l l S[table-format=-3.8]}
\toprule
\multicolumn{1}{c}{$v_1$} & \multicolumn{1}{c}{$v_2$} & \multicolumn{1}{c}{ID$^a$} & \multicolumn{2}{c}{\text{Parameter}} & \text{Value} \\
\midrule

& & & {\color{accent}$\tilde{\nu}_{13} - \tilde{\nu}_{2\times 27}$} & /\si{cm^{-1}}     &        8.719589(3)\\
$2\nu_{27}$     & $\nu_{13}$      &   4000$v_1v_2$ & $G_{b}$    & /\si{\mega\hertz}     &            407.6(1.5)\\
$2\nu_{27}$     & $\nu_{13}$      &   4200$v_1v_2$ & $G_{2b}$   & /\si{\kilo\hertz}     &            -8.720(99)\\
$2\nu_{27}$     & $\nu_{13}$      &   4100$v_1v_2$ & $F_{ac}$   & /\si{\mega\hertz}     &             3.166(25)\\
$2\nu_{27}$     & $\nu_{13}$      &   4101$v_1v_2$ & $F_{ac,J}$  & /\si{\hertz}          &          -120.35(40) \\
& & & {\color{accent} $\tilde{\nu}_{13} - \tilde{\nu}_{26}$} & /\si{cm^{-1}}     &        36.5867(2)\\
$\nu_{26}$      & $\nu_{13}$      &   6100$v_1v_2$ & $F_{ab}$   & /\si{\mega\hertz}     &           6.4435(98) \\
& & & {\color{accent}$\tilde{\nu}_{2\times 27} - \tilde{\nu}_{26}$} & /\si{cm^{-1}}     &        27.8671(2)\\
$\nu_{26}$      & $2\nu_{27}$     &   2100$v_1v_2$ & $F_{bc}$   & /\si{\mega\hertz}     &            2.521(16) \\
\midrule
& & & {\color{darkblue}$\tilde{\nu}_{22} - \tilde{\nu}_{10}$} & /\si{cm^{-1}} &    5.127415(4)\\
$\nu_{10}$      & $\nu_{22}$      &   6000$v_1v_2$ & $G_{c}$    & /\si{\mega\hertz}     &          4080.87(94) \\
$\nu_{10}$      & $\nu_{22}$      &   6200$v_1v_2$ & $G_{2c}$   & /\si{\kilo\hertz}     &          -136.5(7.3) \\
$\nu_{10}$      & $\nu_{22}$      &   6210$v_1v_2$ & $G_{2c,K}$  & /\si{\hertz}          &            2.88(19)  \\
$\nu_{10}$      & $\nu_{22}$      &   6100$v_1v_2$ & $F_{ab}$   & /\si{\mega\hertz}     &             7.35(17) \\
$\nu_{10}$      & $\nu_{22}$      &   6110$v_1v_2$ & $F_{ab,K}$  & /\si{\kilo\hertz}     &           1.424(52)  \\
$\nu_{10}$      & $\nu_{22}$      &   6101$v_1v_2$ & $F_{ab,J}$  & /\si{\hertz}          &             241(15)  \\

\bottomrule
\end{tabular}}
\tnotes{\revchanges{$^a$ The specified IDs are the respective parameter IDs used in the \textit{*.par} and \textit{*.var} files of SPFIT and SPCAT~\cite{Pickett1991, Drouin2017}}.}
\end{threeparttable}
\end{table}

The resulting rotational parameters of the vibrational states are given in \autoref{tab:VibrationalStates} and the interaction parameters are given in \autoref{tab:InteractionParameters}.
In general, the deviations of the rotational constants with respect to the ground vibrational state show good agreement with the rotation-vibration interaction constants from quantum chemical calculations, see \autoref{tab:QCC}.
Typically the obs-calc agreement is within \SI{1}{MHz} but there are some slight deviations of about \SI{2.5}{MHz} for $\alpha^{A}$ of $\nu_{10}$ and strong deviations for $\alpha^{C}$ of $\nu_{10}$ and $\nu_{22}$ of about \SI{\pm 100}{MHz}. 
However, the $\alpha^{C}_{\nu_{10}} + \alpha^{C}_{\nu_{22}}$ values between the experiment and calculations agree well (\revchanges{\SI{15.765(71)}{MHz}} and \SI{15.01}{MHz}, respectively)~\cite{Martin1994,Mills1972}.
As mentioned, these systematic mirror image deviations result from Coriolis coupling contributions to the rotation-vibration interaction constant.
Using equation 13 from Reference~\cite{Mills1972} and the values obtained from the quantum chemical calculations, the Coriolis interaction term of the rotation-vibration interaction constants is calculated to be about \revchanges{$\SI{\pm 103.42}{MHz}$}.
By subtracting this term, the corrected calculated values, 
\revchanges{$\alpha^{C}_{\nu_{10}, \text{corr}} = \SI{-111.18}{MHz} + \SI{103.42}{MHz} = \SI{-7.76}{MHz} $ and $\alpha^{C}_{\nu_{22}, \text{corr}} = \SI{96.17}{MHz} - \SI{103.42}{MHz} = \SI{-7.25}{MHz} $, agree with the experimental values within \SI{1}{MHz}}.

\revchanges{
The two vibrational states closest in energy are the dyad as $\nu_{22}$ is only \SI{5.1}{cm^{-1}} higher in energy than $\nu_{10}$. The strongest interactions are observed for even $\Delta K_a$ values (see \autoref{fig:InteractionInfluenceDyad}) and the interstate transitions have even $\Delta K_a$ and odd $\Delta K_c$ values.
For the triad, $2\nu_{27}$ lies in the middle, \SI{7.9}{cm^{-1}} higher in energy than $\nu_{26}$ and \SI{36.6}{cm^{-1}} lower than $\nu_{13}$ 
}
with the interactions between $2\nu_{27}$ and $\nu_{13}$ being the strongest and appearing for odd $\Delta K_a$ values \revchanges{(see \autoref{fig:InteractionInfluenceTriad})}.
The respective interstate transitions have even $\Delta K_a$ and $\Delta K_c$ values.
The observed trends are similar for both interaction systems, with the center of the interaction increasing in $K_a$ for increasing $J$. %
\revchanges{For the dyad, this pattern starts around $J''=14$ and $K_a'' =3/5$ whereas for $\nu_{13}$ and $2\nu_{27}$ the pattern starts around $J''=31$ and $K_a'' =0/1$.}

For the interaction between $\nu_{10}$ and $\nu_{22}$, only the relative signs of the interaction parameters can be determined (inverting the signs of all interaction parameters results in the same fit/predictions).
For the triad, $\nu_{26}$/$2\nu_{27}$/$\nu_{13}$, there are four equivalent sets of interaction parameter signs (see \ref{sec:InteractionSign}).
Additionally, due to high correlations between the interaction parameters, the presented parameter set should be seen as one of many possible solutions.

\revchanges{
The Coriolis $G_\alpha$ parameters between two fundamentals $r$ and $s$ were estimated from the results of the quantum chemical calculations as follows
\begin{equation}
    G_\alpha(r, s) = \zeta^\alpha_{r, s} * B^\alpha_e * \left( \sqrt{\frac{\omega_{r}}{\omega_{s}}} + \sqrt{\frac{\omega_{s}}{\omega_{r}}} \right)
\end{equation}
Here, $G_\alpha(r, s)$ are the Coriolis parameters of $\alpha = a, b, \text{ or } c$ symmetry, $B^\alpha_e$ the respective equilibrium rotational constant, $\omega$ the harmonic vibrations, and $\zeta^\alpha_{r, s}$ the respective Coriolis coefficients.
As an approximation, we used the experimental ground vibrational state rotational constants for the $B^\alpha_e$ values (see \autoref{tab:isotopologues}) and the harmonic wavenumbers from the quantum chemical calculations (see \autoref{tab:QCC}) for the $\omega$.
For the dyad, this yields $G_c(\nu_{22}, \nu_{10}) = \SI{7854}{MHz}$ which is of same magnitude as the experimental value of \SI{4080.85(14)}{MHz}.
For the triad, only the interaction between the two fundamentals $\nu_{13}$ and $\nu_{26}$ was estimated to be $G_c(\nu_{13}, \nu_{26}) = \SI{875}{MHz}$. However, including $G_c(\nu_{13}, \nu_{26})$ did not improve the fit which hints toward the interaction parameters (especially for the triad) being effective.
}

The resulting \revchanges{weighted root mean square (\textit{WRMS})} values are around \SI{0.8}{} indicating that our uncertainties are slightly conservative.
The analyses of the singly \ce{^{13}C} substituted isotopologues have significantly lower \revchanges{root mean square (\textit{RMS})} and \textit{WRMS} values, which probably results from the lower number of assigned transitions and the limited frequency range (assignments were limited to frequencies below \SI{250}{GHz}).

\revchanges{
\section{Structure determination} 
Using the
improved rotational constants obtained in this work and refitting the line positions of the deuterium substituted isotopologues (\ce{1-\textit{d}1}, \ce{1-\textit{d}2}, \ce{1-\textit{d}3}, 1,2,3,4,\ce{5-\textit{d}5}, and \ce{\textit{d}6}) from Damiani et al.~\cite{Damiani1976}, a semi-experimental equilibrium structure ($r_e^\text{SE}$) has been derived.
Refitting the deuterium isotopic data was essential due to a typo in the $A$ constant of cyclopentadiene 1,2,3,4,\ce{5-\textit{d}5} in Damiani et al.\ (\SI{7707.857(5)}{MHz}~\cite{Damiani1976} vs.\ \SI{7007.857(11)}{MHz} obtained here).
For the deuterated isotopologues, the $A_0$, $B_0$, and $C_0$ rotational parameters were floated while the quartic centrifugal distortion parameters were fixed to their calculated values (fc-CCSD(T)/ANO0 level).
All experimental isotopic $A_0$, $B_0$, and $C_0$ rotational constants were then corrected for the effects of zero-point vibration calculated at the fc-CCSD(T)/ANO0 level (see \autoref{tab:RotParStructureDetermination}) and
the $r_e^\text{SE}$ structure was derived with the STRFIT software \cite{Kisiel2003}. 
A comparison of the $r_e^\text{SE}$ structure against 
\textit{ab initio} values calculated at the fc-CCSD(T)/ANO0 and ae-CCSD(T)/cc-pwCVQZ levels 
is given in \ref{sec:Structures}.
As can be seen, the agreement between the $r_e^\text{SE}$ and the high-level ae-CCSD(T)/cc-pwCVQZ structural parameters is excellent,
to (well) within 10$^{-3}$\,\AA\ and 0.1$^\circ$ for bond lengths and
angles, respectively.
}

\section{Conclusions}
\label{sec:Discussion}
In the present study, the rotational spectrum coverage of cyclopentadiene was extended up to \SI{510}{GHz} by measuring \SI{250}{GHz} of high-resolution broadband spectra.
This allowed to significantly increase the number of assigned lines for the main isotopologue from about 150 to 3510.
As a result, the newly determined parameters produce much more reliable frequency predictions, especially for high frequencies and high quantum numbers (see \autoref{fig:Residuals}).
Especially $R$-branch transitions above \SI{330}{GHz} and $Q$-branch transitions are now in much better agreement.
The presented data allows for astronomical searches at higher frequencies and over a much broader quantum number range.
Similarly, the number of assigned lines for the three singly \ce{^{13}C} substituted isotopologues was increased from about 10 each to well over 200 each.
Their frequency coverage was much improved from only microwave data to up to \SI{250}{GHz}.

Lastly, the dominant vibrational satellite spectra of cyclopentadiene were described for the first time, including their interactions.
Especially the energetically lowest-lying vibrational states are analyzed over a broad quantum number range permitting reliable predictions.

Future work might be targeted at deuterated isotopologues or vibrationally excited states higher in energy.
Whereas analyses of deuterated isotopologues would be simplified by the use of enriched samples, the energetically higher vibrationally excited states could benefit from double-resonance techniques~\cite{Zingsheim2021}.
Additionally, the analyzed vibrational satellite spectrum is a great aid for rovibrational studies in the infrared.
Rovibrational transitions between already analyzed states will be accurately predicted except for the band center and even rovibrational transitions with only one known vibrational state will benefit from the possibility of using techniques like the Automated Spectral Assignment Procedure~\cite{MartinDrumel2015b}.

\section*{Data availability}
The input and output files of SPFIT will be provided as supplementary material.

\section*{Acknowledgments}
\revchanges{We wish to acknowledge an anonymous referee for their careful reading of the manuscript and providing useful suggestions.}
The authors from Cologne gratefully acknowledge the Collaborative Research Center 1601 (SFB 1601 sub-project A4) funded by the Deutsche Forschungsgemeinschaft (DFG, German Research Foundation) – 500700252.

J.-C.G. thanks the national program CNRS PCMI (Physics and Chemistry of the Interstellar Medium) and the CNES for a grant (CMISTEP).
\appendix
\setcounter{figure}{0}
\setcounter{table}{0}

\section{Vibrational wavenumbers and rotation-vibration interaction constants}
\begin{table}[tbh]
\caption{Rotation-vibration interaction constants $\alpha_i$, \revchanges{harmonic and anharmonic wavenumbers} $\tilde{\nu_i}$ from quantum chemical calculations at the CCSD(T)/ANO0 level as well as the wavenumbers reported by Castellucci et al.~\cite{Castellucci1975}. Rotation-vibration interaction constants are in \si{MHz} and wavenumbers in \si{cm^{-1}}. Literature values are taken from liquid phase spectra if not specified otherwise.}
\label{tab:QCC}
\resizebox{\linewidth}{!}{
\begin{tabular}{l l S[table-format=-2.3] S[table-format=-2.3] S[table-format=-2.3] S[table-format=3.0] S[table-format=3.0] S[table-format=3.0]}
        \toprule
        \multicolumn{2}{l}{Mode} & \text{$-\alpha^{A}_\text{calc}$} & \text{$-\alpha^{B}_\text{calc}$} &  \text{$-\alpha^{C}_\text{calc}$}  &  \text{$\tilde{\nu}_\text{calc}^\text{anh}$}  & \text{$\tilde{\nu}_\text{calc}^\text{harm}$}  &  \text{$\tilde{\nu}_\text{lit}$}~\cite{Castellucci1975} \\

        \midrule
        $\nu_{27}  $ & $B_1$ &     3.77 &   -11.22 &     2.93 &   339 &    340 &   350 \\
        $\nu_{14}  $ & $A_2$ &    -6.86 &   -17.19 &     2.74 &   497 &    502 &   516$^a$ \\
        $\nu_{26}  $ & $B_1$ &   -16.56 &    -8.26 &     1.15 &   658 &    670 &   664$^b$ \\
        $\nu_{13}  $ & $A_2$ &   -30.06 &    -4.22 &     1.94 &   691 &    701 &   700 \\
        $\nu_{10}  $ & $A_1$ &    -6.44 &     2.62 &  -111.18 &   789 &    799 &   802 \\
        $\nu_{22}  $ & $B_2$ &    18.45 &     2.51 &    96.17 &   793 &    804 &   805 \\
        $\nu_{25}  $ & $B_1$ &  -830.70 &   -33.86 &    -0.80 &   896 &    920 &   891 \\
        $\nu_{12}  $ & $A_2$ &   -19.44 &   -55.71 &    -0.81 &   897 &    917 &   941$^a$ \\
        $\nu_{9}   $ & $A_1$ &   815.34 &    27.89 &   -20.22 &   903 &    920 &   915 \\
        $\nu_{24}  $ & $B_1$ &     2.38 &   -12.17 &     0.22 &   905 &    927 &   925 \\
        $\nu_{21}  $ & $B_2$ &   -11.03 &    23.84 &     5.81 &   953 &    972 &   959 \\
        $\nu_{8}   $ & $A_1$ &     2.39 &   -14.20 &    -4.82 &  1015 &   1006 &   994 \\
        $\nu_{20}  $ & $B_2$ &    17.40 &     7.20 &    -4.29 &  1085 &   1104 &  1090 \\
        $\nu_{7}   $ & $A_1$ &    14.67 &    -1.33 &     2.07 &  1102 &   1120 &  1106 \\
        $\nu_{11}  $ & $A_2$ &    -8.06 &    -0.53 &     2.43 &  1102 &   1131 &  1100$^a$ \\
        $\nu_{19}  $ & $B_2$ &    -0.34 &     1.60 &    -9.04 &  1236 &   1266 &  1239 \\
        $\nu_{18}  $ & $B_2$ &    -1.90 &     0.52 &    -8.39 &  1278 &   1310 &  1292 \\
        $\nu_{6}   $ & $A_1$ &     2.59 &    -3.70 &    -5.98 &  1360 &   1393 &  1365 \\
        $\nu_{5}   $ & $A_1$ &    -4.14 &     9.97 &     7.68 &  1389 &   1428 &  1378 \\
        $\nu_{4}   $ & $A_1$ &   -17.62 &    -4.69 &    -5.98 &  1503 &   1547 &  1500 \\
        $\nu_{17}  $ & $B_2$ &   -15.81 &    -3.48 &    -4.02 &  1572 &   1615 &  1580 \\
        $\nu_{3}   $ & $A_1$ &    -4.70 &    -4.77 &    -0.72 &  2905 &   3042 &  2886 \\
        $\nu_{23}  $ & $B_1$ &    -3.34 &    -3.42 &    -1.21 &  2929 &   3086 &  2900 \\
        $\nu_{16}  $ & $B_2$ &    -6.12 &    -7.20 &    -3.40 &  3084 &   3216 &  3043 \\
        $\nu_{2}   $ & $A_1$ &    -6.23 &    -7.71 &    -3.53 &  3090 &   3225 &  3075 \\
        $\nu_{1}   $ & $A_1$ &    -6.84 &    -7.91 &    -3.67 &  3113 &   3250 &  3091 \\
        $\nu_{15}  $ & $B_2$ &    -6.86 &    -7.52 &    -3.53 &  3115 &   3242 &  3105 \\
        \bottomrule
\end{tabular}}
\tnotes{
$^{a}$~Solid-phase value (other literature energy values are taken from liquid phase spectra).\\
$^{b}$~Gas-phase high-resolution infrared spectroscopy yields a value of 663.84800(5)\,cm$^{-1}$ \cite{Boardman1990}.
}

\end{table}

\section{Rotational constants from literature and quantum chemical calculations}
\begin{table}[tbh]
\caption{Literature values of the rotational constants for the main isotopologue (from the CDMS~\cite{Endres2016, Mller2001}) and for the singly substituted \ce{^{13}C} isotopologues (from Damiani et al.~\cite{Damiani1976})}
\label{tab:LiteratureData}
\centering
\begin{threeparttable}
\resizebox{\linewidth}{!}{
\begin{tabular}{l l S[table-format=5.6] S[table-format=5.4] S[table-format=5.4] S[table-format=5.6]}
        \toprule
        \multicolumn{2}{l}{\text{Param.}} & \ce{$c$-C5H6}~\cite{Endres2016, Mller2001}     &   \ce{1-^{13}C}~\cite{Damiani1976} & \ce{2-^{13}C}~\cite{Damiani1976} & \ce{3-^{13}C}~\cite{Damiani1976} \\
        \midrule
        $A$ & /\si{MHz} & 8426.1111(11)    &   8226.028(9)  &   8419.949(6)  &  8345.097(10)  \\
        $B$ & /\si{MHz} & 8225.6387(11)    &   8219.434(7)  &   8040.358(6)  &  8108.662(7)   \\
        $C$ & /\si{MHz} & 4271.4377(11)    &   4217.760(6)  &   4219.419(4)  &  4219.067(7)  \\
     $-D_J$ & /\si{kHz} &   -2.6948(13)    &   \text{--} &  \text{--} & \text{--} \\
  $-D_{JK}$ & /\si{kHz} &    4.0626(29)    &   \text{--} &  \text{--} & \text{--} \\
   $-D_{K}$ & /\si{kHz} &   -1.6842(17)    &   \text{--} &  \text{--} & \text{--} \\
      $d_1$ & /\si{Hz}  &  -39.7(18)       &   \text{--} &  \text{--} & \text{--} \\
      \midrule
      \multicolumn{2}{l}{Transitions}   & 157     &   \text{--} &  \text{--} & \text{--} \\
      \multicolumn{2}{l}{Lines}         & 92      &   \text{--} &  \text{--} & \text{--} \\
      \textit{RMS} &/\si{\kilo\hertz}   & 30.68  &   \text{--} &  \text{--} & \text{--} \\
      \multicolumn{2}{l}{\textit{WRMS}} & 1.02 &   \text{--} &  \text{--} & \text{--} \\
        \bottomrule
    \end{tabular}}
\end{threeparttable}
\end{table}

\begin{table}[tbh]
\caption{\revchanges{Quartic centrifugal distortion constants ($\text{III}^\text{l}$ representation) of the main isotopologue and the singly substituted \ce{^{13}C} isotopologues as calculated at the fc-CCSD(T)/ANO0 level.}}
\label{tab:QCCQuartics}
\centering
\begin{threeparttable}
\resizebox{\linewidth}{!}{
\begin{tabular}{l l S[table-format=4.6] S[table-format=4.6] S[table-format=4.6] S[table-format=4.6] S[table-format=5.6]}
        \toprule
        \multicolumn{2}{l}{\text{Param.}}  & \ce{$c$-C5H6}  & \ce{1-^{13}C} & \ce{2-^{13}C} & \ce{3-^{13}C} \\
        \midrule
     $-D_J$ & /\si{kHz} & -2.58744   &  -2.54647      &  -2.52934      & -2.5328  \\
  $-D_{JK}$ & /\si{kHz} &  3.89513   &   3.83687      &   3.80924      &  3.81731 \\
   $-D_{K}$ & /\si{kHz} & -1.61171   &  -1.58884      &  -1.57739      & -1.58123 \\
      $d_1$ & /\si{Hz}  & -44.0093   &   61.0209      &  -20.1082      & -11.8137 \\
      $d_2$ & /\si{Hz}  &  0.751250  &   -2.71914     &  -3.62988      &  1.22695 \\
        \bottomrule
    \end{tabular}}
\end{threeparttable}
\end{table}

\section{Interaction hotspots}
\begin{figure}[tbh]
    \centering
    \includegraphics[width=\linewidth]{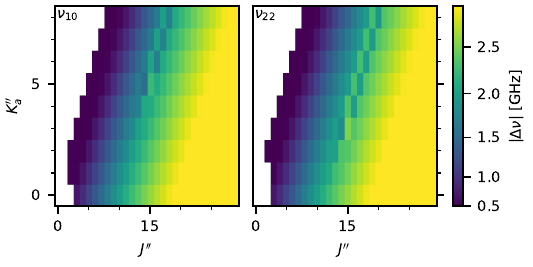}
    \caption{Absolute shifts in transition frequency between predictions with and without interactions for $\Delta J_{\Delta K_a, \Delta K_c} = 1_{1, 1}$ transitions.
    The heatmaps of the interacting vibrational states $\nu_{10}$ (left) and $\nu_{22}$ (right) show the same pattern when shifted by $\Delta K_a = 2$.
    The center of the interactions increases in $K_a$ with increasing $J$ value.}
    \label{fig:InteractionInfluenceDyad}
\end{figure}

\begin{figure}[tbh]
    \centering
    \includegraphics[width=\linewidth]{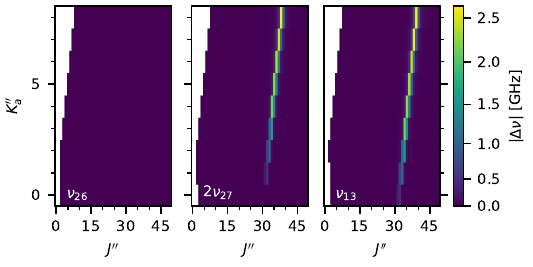}
    \caption{\revchanges{Absolute shifts in transition frequency between predictions with and without interactions for $\Delta J_{\Delta K_a, \Delta K_c} = 1_{1, 1}$ transitions.
    The heatmaps of the interacting vibrational states $\nu_{26}$ (left), $2\nu_{27}$ (middle) and $\nu_{13}$ (right) highlight that the strongest interactions are between $2\nu_{27}$ and $\nu_{13}$ with $\Delta K_a = 2$.
    The center of the interactions increases in $K_a$ with increasing $J$ value.
    In contrast to \autoref{fig:InteractionInfluenceDyad}, the $x$-axis and the $z$-axis (being the heatmap color) are scaled differently to highlight the main interactions better.}}
    \label{fig:InteractionInfluenceTriad}
\end{figure}

\section{Semi-experimental equilibrium structure}
\label{sec:Structures}

\begin{table}[tbh]
	\caption{\revchanges{Rotational constants (MHz) and zero-point vibrational corrections (fc-CCSD(T)/ANO0, MHz)  of cyclopentadiene isotopologues.}}
	\label{tab:RotParStructureDetermination}
	\centering
	\begin{threeparttable}
	\resizebox{\linewidth}{!}{
	\begin{tabular}{l S[table-format=4.3] S[table-format=4.3] S[table-format=4.3] S[table-format=4.3] S[table-format=4.3] S[table-format=4.3] S[table-format=4.3] S[table-format=4.3] S[table-format=4.3] }
			\toprule
			 & \ce{$c$-C5H6}  & \ce{1-^{13}C1} & \ce{2-^{13}C} & \ce{3-^{13}C} & 	\ce{1-\textit{d}1} & \ce{1-\textit{d}2} & \ce{1-\textit{d}3} & \text{1,2,3,4,\ce{5-\textit{d}5}} & \ce{\textit{d}6} \\
			\midrule
$A_0             $ & 8426.109 & 8226.053 & 8420.043 & 8345.132 & 8129.952 & 8414.045 & 8307.036 & 7007.858 & 6608.398 \\
$B_0             $ & 8225.640 & 8219.483 & 8040.425 & 8108.714 & 7859.534 & 7591.889 & 7678.126 & 6681.888 & 6607.340 \\
$C_0             $ & 4271.437 & 4217.759 & 4219.411 & 4219.065 & 4145.090 & 4091.154 & 4090.254 & 3529.302 & 3444.129 \\
$\Delta A        $ &   63.046 &   66.046 & 62.748   &   63.249 &   65.660 &   63.130 &   64.793 &   49.690 &   48.552 \\
$\Delta B        $ &   66.461 &   60.796 & 64.627   &   63.844 &   58.864 &   58.628 &   56.982 &   48.475 &   46.425 \\
$\Delta C        $ &   34.228 &   33.512 & 33.672   &   33.580 &   33.263 &   32.018 &   31.969 &   25.954 &   25.158 \\
$A_e^{\text{SE}} $ & 8489.155 & 8292.100 & 8482.792 & 8408.381 & 8195.612 & 8477.175 & 8371.829 & 7057.549 & 6656.950 \\
$B_e^{\text{SE}} $ & 8292.102 & 8280.279 & 8105.053 & 8172.558 & 7918.399 & 7650.517 & 7735.108 & 6730.363 & 6653.765 \\
$C_e^{\text{SE}} $ & 4305.665 & 4251.271 & 4253.082 & 4252.645 & 4178.353 & 4123.172 & 4122.222 & 3555.255 & 3469.288 \\
			\bottomrule
		\end{tabular}}
	\end{threeparttable}
	\end{table}

\revchanges{
Equilibrium structural parameters of cyclopentadiene (bond lengths in \AA, angles in degrees) calculated at the fc-CCSD(T)/ANO0 and ae-CCSD(T)/cc-pwCVQZ levels of theory and determined semi-experimentally. See text for details.}

\begin{small}
\begin{verbatim}
X
X 1 rd
C 2 rd 1 a90
C 3 r1 2 a1 1 d90
C 3 r1 2 a1 4 d180
C 4 r2 3 a2 5 d0
C 5 r2 3 a2 4 d0
H 4 r3 3 a3 2 d180
H 5 r3 3 a3 2 d180
H 6 r4 4 a4 8 d0
H 7 r4 5 a4 9 d0
H 3 r5 2 a5 1 d0
H 3 r5 2 a5 12 d180

           ANO0     /   pwCVQZ   /     reSE       
rd   =    1.0       /   1.0      /   1.0
a90  =   90.0       /  90.0      /  90.0 
r1   =    1.515110  /   1.500253 /   1.49906(34)
a1   =   51.526426  /  51.537508 /  51.541(15)
d90  =   90.0       /  90.0      /  90.0
d180 =  180.0       / 180.0      / 180.0
r2   =    1.361296  /   1.346357 /   1.34635(35)
a2   =  109.368447  / 109.329183 / 109.330(24)
d0   =    0.0       /   0.0      /   0.0
r3   =    1.087290  /   1.078491 /   1.07865(24)
a3   =  124.008184  / 124.075312 / 124.01(11)
r4   =    1.087975  /   1.079202 /   1.07907(22)
a4   =  126.235302  / 126.127498 / 126.082(60)
r5   =    1.102197  /   1.093823 /   1.09441(17)
a5   =  126.576616  / 126.701234 / 126.710(12)
\end{verbatim}
\end{small}

\section{Comparison of initial and final residuals}
\begin{figure}[tbh]
    \centering
    \includegraphics[width=\linewidth]{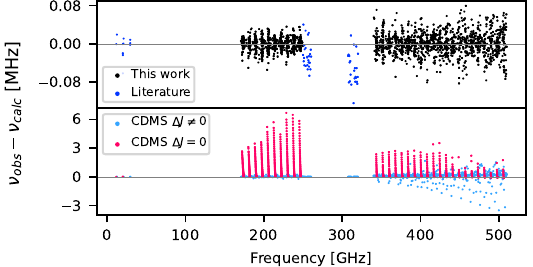}
    \caption{Residuals of the ground vibrational state of the main isotopologue for the predictions from this work (top) compared with the residuals for the predictions from the CDMS~\cite{Endres2016, Mller2001} (based on data of previous reports~\cite{Bogey1988,Scharpen1965}). Colors distinguish between literature assignments and new assignments in the top row and between $Q$- and $R$-branch transitions in the bottom row. This work greatly improves predictions for $R$-branch transitions above \SI{330}{GHz} and $Q$-branch transitions above \SI{40}{GHz}.}
    \label{fig:Residuals}
\end{figure}

\section{Signs of Coriolis interaction parameters}
\label{sec:InteractionSign}

It is common when treating Coriolis interactions that there are multiple equivalent parameter sets for the analysis which only differ in the signs of the Coriolis interaction parameters~\cite{Islami1996,Mller2023}.
Even though the calculated Coriolis interaction parameters have a defined sign, it typically cannot be determined experimentally.
The equivalent interaction parameter sets of the $\nu_{10}$/$\nu_{22}$ diad are
\revchanges{
\begin{equation}
    \begin{aligned}
        &\{ G_{c}, G_{2c}, G_{2c,K}, F_{ab}, F_{ab, K}, F_{ab, J} \} \\
        &\{-G_{c},-G_{2c},-G_{2c,K},-F_{ab},-F_{ab, K},-F_{ab, J} \} \\
    \end{aligned}
\end{equation}
}
This means inverting all parameter signs results in exactly the same fit/frequency predictions.
The triad, being $\nu_{26}$, $2\nu_{27}$, and $\nu_{13}$, has four equivalent interaction parameter sets, being 
\revchanges{
\begin{equation}
    \begin{aligned}
        &\{ G_b^{12}, G_{2b}^{12}, F_{ac}^{12}, F_{ac, J}^{12}, F_{bc}^{01}, F_{ab}^{02} \} \\
        &\{ G_b^{12}, G_{2b}^{12}, F_{ac}^{12}, F_{ac, J}^{12},-F_{bc}^{01},-F_{ab}^{02} \} \\
        &\{-G_b^{12},-G_{2b}^{12},-F_{ac}^{12},-F_{ac, J}^{12},-F_{bc}^{01}, F_{ab}^{02} \} \\
        &\{-G_b^{12},-G_{2b}^{12},-F_{ac}^{12},-F_{ac, J}^{12}, F_{bc}^{01},-F_{ab}^{02} \} \\
    \end{aligned}
\end{equation}}
where for brevity the superscript indices specifying the vibrational states are given as their respective vibrational identifiers used in the \textit{*.par} file of SPFIT: $0 \equiv \nu_{26}$,  $1 \equiv \nu_{13}$, and $2 \equiv 2\nu_{27}$.

For both interacting systems, the relative signs within one Coriolis symmetry and two-state connection have to stay fixed.
For the triad, the signs for the different Coriolis symmetries can be exchanged in pairs.

\biboptions{comma,sort&compress}
\bibliographystyle{elsarticle-num}
\bibliography{resources/bibliography}

\end{document}